\documentclass[jap,aip,preprint]{revtex4-1}

\usepackage{graphicx}  % needed for figures
\usepackage{dcolumn}   % needed for some tables
\usepackage{bm}        % for math
\usepackage{amssymb}
\usepackage[english]{babel}
\usepackage{lineno}
\usepackage{hyperref}
\usepackage{relsize}
\begin{document}

%\preprint{APS/123-QED}

\title{Anomalous compressibility in 1T$^\prime$ MoTe$_{2}$ single crystal:High pressure Raman and structural studies}

\author{Bishnupada Ghosh}
\affiliation{Department of Physical Sciences, Indian Institute of Science Education and Research Kolkata, Mohanpur Campus, Mohanpur 741246, Nadia, West Bengal, India.}

\author{Pinku Saha}
\affiliation{Department of Physical Sciences, Indian Institute of Science Education and Research Kolkata, Mohanpur Campus, Mohanpur 741246, Nadia, West Bengal, India.}

\author{Goutam Dev Mukherjee}
\email [Corresponding author:]{goutamdev@iiserkol.ac.in}
\affiliation{Department of Physical Sciences, Indian Institute of Science Education and Research Kolkata, Mohanpur Campus, Mohanpur 741246, Nadia, West Bengal, India.}

\begin{abstract}
	A detailed high pressure study is carried out on 1T$^\prime$ MoTe$_{2}$ using X-ray diffraction(XRD) and Raman spectroscopy measurements upto about 30.5 GPa. High pressure XRD measurements show no structural transition. All the lattice parameters exhibit anomalous changes in the pressure region 8.4 to 12.7 GPa. Compressibility of the sample is found to be reduced by almost four times above 12.7 GPa with respect to that below 8.4 GPa. The anomalies in the Raman mode corresponding to the out of plane vibrations of Mo atoms sitting in the unit cell surface indicate a strong electron phonon coupling possibly mediated by differential strain inside the unit cell. 
\end{abstract}

\date{\today}
\maketitle

%\tableofcontents

\section{\label{sec:level1}Introduction}

Transition metal dichalcogenides(TMDCs) have been in research forefront due to their variable band gap and strong spin orbit coupling\cite{Qingjun}. MoTe$_{2}$ is one of the recently discovered highly promising 2D TMD synthesized in different crystal structures: hexagonal(2H), monoclinic(1T$^\prime$) and orthorhombic(T$_{d}$) phases\cite{Xiaoli}. 2H MoTe$_{2}$ is a semiconductor; 1T$^\prime$ and T$_{d}$ are semi-metals. 1T$^\prime$ MoTe$_{2}$ exhibits superconductivity at very low temperature and the transition temperature is found to be enhanced by application of pressure\cite{Yanpeng}.
High pressure studies have revealed a strain mediated change in crystal structure of exfoliated a few layered WS$_{2}$ and MoSe$_{2}$ samples\cite{Pinku,Pinku2}. A recent study on 2H MoTe$_{2}$ has shown a semiconducting to metallic transition above 10 GPa without any structural transition\cite{Zhao}. Therefore a detailed high pressure investigation on 1T$^\prime$ MoTe$_{2}$ is necessary to understand its strain and electronic structure correlation.

In the present work we have carried out a detailed high pressure investigation on monoclinic MoTe$_{2}$ using Raman spectroscopy and X-ray diffraction measurements. Our results show anomalies in equation of state parameters and certain Raman modes from 8.4 GPa to 12.7 GPa indicating a strain mediated electronic transition.

\section{Experiment }
High pressure XRD and Raman spectroscopy experiments are carried out using piston-cylinder type diamond-anvil cell (DAC) from Almax easyLab\cite{Pinku}. The culets of the diamonds are of 300 $\mu m  $ diameter. Sample is loaded inside a central hole of diameter 100 $\mu m$ drilled on a steel gasket preindented to a thickness of 45 $\mu m  $. For Raman spectroscopy measurements small ruby chips of (approximate size 3-5 $\mu m  $) are loaded along with the sample inside the hole to determine the pressure\cite{Mao}. High quality single crystal 1T$^\prime$  MoTe$_{2}$ sample is purchased from HQ graphene. Cobolt-samba diode pump laser of wavelength 532 nm and Sapphire SF optically pumped semiconductor laser of wavelength 488 nm are used as excitation sources for Raman measurements. Raman spectra are collected in the back scattering geometry using a confocal micro-Raman system (Monovista from SI GmbH) with 750mm monochromator and a back-illuminated PIXIS 100BR (1340X100) CCD camera. The scattered light is collected with 50x infinitely corrected long working distance objective with numerical aperture 0.42. The collected scattered light is dispersed using a grating with 1500 g/mm with a spectral resolution better than 1.2 cm$^{-1}$. 
High pressure XRD study is carried out at XPRESS beam-line in ELETTRA synchrotron source on bulk  MoTe$_{2}$  using the diamond anvil cell described above. A mixture of Methanol-Ethanol (4:1) is used as liquid pressure transmitting medium. Very small amount of silver powder is mixed with the sample to measure the pressure inside the diamond anvil cell. Monochromatic X-rays of wavelength 0.4957 \AA\ is used. The incident X-ray is collimated to 20 $\mu m$ using a pin hole. Diffracted XRD patterns are collected using MAR 345 image plate detector system aligned normal to the beam. XRD pattern of LaB$_{6}$ is used to determine sample to detector distance. 2D XRD images are converted into 2$\theta$ vs intensity  using Dioptas software\cite{Preschera}. Pressure is calibrated using 3rd order Birch Murnaghan equation of state of silver\cite{Dewaele}. Pressure is varied in small steps manually up to 30.5 GPa, the highest pressure of this XRD study. X-ray diffraction patterns are indexed using CRYSFIRE software\cite{Shirley} and Lebail fittings are performed using GSAS software\cite{Toby}.

\section{Results and Discussions}
%%\subsection{X-ray diffraction studies}

Fig.1(b) shows the ambient XRD pattern of MoTe$_{2}$. Indexing the XRD pattern reveals monoclinic crystal structure (1T$^\prime$) having P2$_{1}$/m space group symmetry with lattice parameters, $a=$6.3279(2) \AA , $b=$3.4747(1) \AA , $c=$13.8605(1) \AA, $\beta=$ 93.624(8)$^\circ$  and volume $V$=304.15(4) \AA$^{3}$ and $Z$=2, which matches well with literature \cite{Dawson}.  %(pressure evolution of XRD patterns are shown in supplementary information).
In Fig.1(a) we have shown the pressure evolution of XRD patterns till about 31 GPa, the highest pressure of this study. No change in crystal structure is observed in XRD patterns. Large broadening of the XRD patterns are observed above 12 GPa, may be due to increase strain. However with pressure release the sample comes back to its original state.
 %In Fig.1 we have shown Rietveld refinements of XRD patterns at ambient and Lebail fitting for selected high pressure points. 
Relative lattice parameter values are shown in Fig.2(a). It is found that the $c$-axis is more compressible at low pressures compared to other two axes. Applying pressure upto 30.5 GPa it is observed that $a$ and $b$ axes are reduced by about 5\% and 6\%, respectively; where as $c$ axis is reduced by more than 12\%. Close inspection of Fig.2a shows that $a$ and $b$ axes decrease almost linearly up to about 8.4 GPa, and then show a plateau type characteristic up to about 12.7 GPa followed by an almost linear decrease. $c$ axis decreases significantly (about 5\%) up to about 3.9 GPa and then decreases slowly with similar plateau type behavior in the pressure range 8.4 GPa to 12.7 GPa. Variation of monoclinic unit cell angle $\beta$ with pressure is shown in Fig.2b, which shows a peak at about 10.4 GPa. From all the above data it may be noted that the lattice shows an anomalous compression behavior in the pressure range 8.4 GPa to 12.7 GPa. Variation of unit cell volume with applied pressure is shown in Fig.2(c). Volume data cannot be fitted using a single equation of state(EOS). The unit cell volume initially decreases by about 12\% upto pressure 8.4 GPa and could only be fitted using $4^{th}$ order BM-EOS\cite{Ahmad}. 
\begin{equation}
P=\frac{9}{16}B_{0}[-B_{1}x^{-\frac{5}{3}}+B_{2}x^{-\frac{7}{3}}-B_{3}x^{-\frac{5}{3}}+B_{4}x^{-\frac{11}{3}}],
\end{equation}
\noindent
[ where $B_{1}=B_{0}B^{''}+(B^{'}-4)(B^{'}-5)+\frac{59}{9},
B_{2}=3B_{0}B^{''}+(B^{'}-4)(3B^{'}-13)+\frac{129}{9},\\
B_{3}=3B_{0}B^{''}+(B^{'}-4)(3B^{'}-11)+\frac{105}{9},
B_{4}=B_{0}B^{''}+(B^{'}-4)(B^{'}-3)+\frac{35}{9}$]\\
\noindent
The Bulk modulus $B_{0}$ and its first pressure derivative $B^\prime$ are found to be 24$\pm$2 GPa and 15$\pm$1, respectively, which show that MoTe$_{2}$ is highly compressible at low pressures. Higher pressure volume data could be fitted well with $3^{rd}$ order BM-EOS above 12.7 GPa\cite{Angel}.
\begin{equation}P=\frac{3B_{0}}{2}[(\frac{V_{0}}{V})^{\frac{7}{3}}-(\frac{V_{0}}{V})^{\frac{5}{3}}][1+\frac{3}{4}(B^{'}-4)[(\frac{V_{0}}{V})^{\frac{2}{3}}-1]]
\end{equation}
\noindent
 $B_{0}$ and $B^{\prime}$ for this region are 98$\pm$2 GPa and 3.0$\pm$0.2, respectively. Such large change in compressibility in between two pressure ranges without any change in structure, shows that the internal strain of the lattice changes anomalously in the pressure range 8.4-12.7 GPa. Anomalies in lattice parameters and EOS observed in the pressure range (8.4-12.7 GPa) are close to the freezing pressure of PTM\cite{Klotz}. Therefore there is always a possibility that the observed anomalies may be related to non-hydrostatic stress induced by freezing of PTM. However, the large non-linear compression behaviour in $c$ axis is observed at much lower pressures up to about 3.9 GPa. Similarly $a$,$b$ lattice parameters show a step like characteristic in the range 8.4-12.7 GPa pressure. One would expect that effect of non hydrostatic stress should have resulted in just a slope change or a shift in data just above 10 GPa, the freezing pressure of PTM\cite{Klotz}, not a change in compressibility behaviour as observed.
In addition our experimental geometry contained much less sample in comparison to PTM volume and XRD patterns were collected from the central 20  $\mu m  $ area. Therefore the effect of non-hydrostatic stress from freezing is expected to be much less. 

%%\subsection{Raman Spectroscopy Studies}

For complementary investigation of the sample we have carried out Raman spectroscopy measurements at high pressures. Monoclinic structure of 1T$^\prime$ MoTe$_{2}$ with a space group P2$_{1}$/m allows 18 Raman active phonon modes (12 A$_{g}$ + 6 B$_{g}$) \cite{Xiaoli}. At room temperature and ambient pressure in back scattering geometry, 9 A$_{g}$ and 3 B$_{g}$ modes are detected in our experiment, which match well with literature\cite{Xiaoli}. Raman spectra at ambient conditions are collected using 532 nm and 488 nm laser sources to see effect of  excitation energies if any and are shown in Fig.3. In both the cases the spectral features appear more or less similar. The Raman spectra in both cases are fitted using a Lorentzian and the obtained mode frequencies values are given in Table-1. We have carried out high pressure Raman Spectroscopy measurements using two PTMs: (4:1) Methanol-Ethanol mixture and Silicone oil. First we shall discuss the results of the experiments using (4:1)Methanol-Ethanol mixture. 
The pressure evolution of Raman spectra using both pressure transmitting media are shown in Fig.4. Among all the detected modes, $^{8}$A$_{g}$(162 cm$^{-1}$) and $^{12}$A$_{g}$ (258 cm$^{-1}$) modes remain prominent at high pressures. Intensity of $^{2}$B$_{g}$ mode at 93 cm$^{-1}$ is found to become prominent above 9.7 GPa, which is not visible properly at lower pressure points. Pressure evolution of a few selected mode frequencies are shown in Fig.5(a). We find a sudden red shift in $^{12}A_{g}$ mode around 12 GPa as shown in Fig.6(a). However that may have been induced by freezing of PTM. Most of the modes show highly non-linear behavior at lower pressures and can be related to the large anomalous compressibility of the sample(particularly along $c$ axis). Our results are in contrast to high pressure Raman study carried out by Qi \textit{et al}\cite{Yanpeng} showing no change in slope of pressure evolution of Raman mode frequency. To check whether the anomalies observed in our data are due to experimental artifacts we have carried out high pressure experiments with Silicone oil pressure transmitting medium and are shown in Fig.4(b). The spectra are found to be similar to previous experiments using Methanol-Ethanol mixture. The variation of few selected Raman mode frequencies with pressure for Silicone oil are shown in Fig.5(b). The Raman modes corresponding to $^{8}$A$_{g}$, $^{12}$A$_{g}$ and  $^{6}$A$_{g}$  show a slope change at about 8.5 GPa. We also observe a slight red-shift in $^{12}$A$_{g}$ (Fig.6(b)), however the shift is smaller in this case in comparison to Methanol-Ethanol PTM (within the error limits). Also the slope changes in $^{8}$A$_{g}$ and $^{12}$A$_{g}$ modes, which were masked due to freezing of Methanol-Ethanol PTM, becomes apparent in case of Silicone oil PTM. This indicates the results observed can be attributed to sample properties. Change of FWHM for $^{12}$A$_{g}$ mode is shown in Fig.6(c-d) for both Methanol- Ethanol mixture and  Silicone oil. It remains almost constant up to about 12 GPa, followed by a sharp rise  up to about 20 GPa. Sudden increase in FWHM of the above Raman mode, which has shown a red-shift, is due to decrease in life-time of the above phonon mode. Life time of a phonon mode is affected due to its scattering with other phonons or electrons. Since we do not see any change in structure, the compression is resisted by the valence electron clouds of Mo and Te atoms taking part in the bond formation. Therefore this sudden increase of FWHM of this mode then can be attributed to electron-phonon scattering due to excited electronic states. $^{12}$A$_{g}$ mode corresponds to the out-of plane vibration of Mo-atoms sitting at the unit cell surface in the $ab$-plane along c-axis\cite{Xiaoli}. Therefore anomalous large compression of $c$-axis probably leads to the observed anomalies in  $^{12}$A$_{g}$ mode. Zhao et al\cite{Zhao} have shown a gradual electronic phase transition from semiconducting to metallic state in 2H MoTe$_{2}$ at 9.6 GPa. They have seen a slope change in Raman frequency variation with pressure above 9.6 GPa. 
Anomalies in Raman mode parameters as discussed above are observed in the same pressure range, where anomalies in the unit cell lattice parameters and compressibilities are observed from XRD measurements. Changes in compressibility behavior below and above pressure range 8.4 to 12.7 GPa, indicate to changes in internal arrangements of atoms inside unit cell leading to differential strain. 

%In Fig.7 we have compared the configuration of atoms in unit cell of 1T$^\prime$ at ambient,6 GPa, 9.5 %GPa, 12 GPa ,15.9 GPa and 30.5 GPa. Ambient 1T$^{\prime}$ structure shows clearly 3-layers of Mo-Te %chains. With increasing pressure it is obvious from the plots that configuration of Mo-Te atoms in %1$^{st}$ and 3$^{rd}$ layers remains almost intact. The pressure compression is accommodated by %rotation of Mo-Te atomic configuration in the middle layers. This change in internal structure of the %unit cell leads to the anomalous behavior of observed parameters in XRD and Raman studies.

 At the present situation it is difficult to comment on exact change in electronic behavior of the sample, though a sudden increase in FWHM of $^{12}$A$_{g}$ Raman mode at 12 GPa does point to an electron-phonon coupling effect. Qi \textit{et al} \cite{Yanpeng} have shown that superconducting transition temperature of 1T$^{\prime}$  MoTe$_{2}$ increases with pressure with a peak at 11.7 GPa. High pressure electronic structure calculations using by the same group show a large increase in electron DOS at Fermi level upto 12 GPa. Therefore one may conclude that the drastic change in electronic DOS at lower pressure are probably due to large compressibility as observed from our experiments.
  This can be possibly used to predict an electronic transition in 1T$^{\prime}$ MoTe$_{2}$ in the pressure range 8.4 GPa to 12.7 GPa. However one needs to carry out other experimental and theoretical studies under pressure to verify this.

\section{Conclusion}
We have carried out detailed high pressure Raman and XRD investigation to probe vibrational and structural properties of high quality single crystal 1T$^{\prime}$ MoTe$_{2}$. High pressure XRD analysis shows a drastic change in compressibility of 1T$^{\prime}$  MoTe$_{2}$ single crystal in between pressure ranges, ambient to 8.4 GPa and above 12.7 GPa, however no structural transition is observed.  Anomalies found in few Raman mode frequencies, red shift of $^{12}$A$_{g}$ and drastic change in FWHM of $^{12}$A$_{g}$ mode are found to happen around 12 GPa. All these changes found in Raman measurements can be interpreted in terms of the large change of compressibility inducing differential strain inside the unit cell which can lead to an electronic transition. These studies will help in understanding the field of layered TMDCs having superconducting behavior and may lead to prediction of novel strain modulated optoelectronic devices.

{\bf Acknowledgments}
The authors gratefully acknowledge the Ministry of Earth Sciences, Government of India, for the financial support under the grant No. MoES/16/25/10-RDEAS to carry out this high pressure research work. B.Ghosh and P.Saha also gratefully acknowledge Department of Science and Technology, Government of India for their INSPIRE fellowship grant for pursuing PhD program. The authors also gratefully acknowledge the financial support from the Department of Science and Technology, Government of India to visit XPRESS beamline in the ELETTRA Synchrotron light source under the Indo-Italian Executive Programme of Scientific and Technological Cooperation.

\newpage
\begin{table}
\caption{\label{tab:table1}Ambient Raman mode frequencies of MoTe$_{2}$  }
\begin{ruledtabular}
\begin{tabular}{ccc}
Raman Modes & Using 488 nm laser & Using 532 nm laser \\
\hline
	$^{1}$A$_{g}$   & 76    & 77   \\
	$^{2}$A$_{g}$   & 83    & 83  \\
	$^{2}$B$_{g}$   & 93    & 94  \\
	$^{3}$B$_{g}$   & 106   & 106   \\
	$^{3}$A$_{g}$   & 110   & 110  \\
	$^{5}$A$_{g}$   & 122   & 122  \\
	$^{6}$A$_{g}$   & 127   & 128  \\
	$^{7}$A$_{g}$   & 154   & 155  \\
	$^{8}$A$_{g}$   & 162   & 162   \\
	$^{6}$B$_{g}$   & 190   & 191  \\
	$^{9}$A$_{g}$   & 246   & 247 \\
	$^{12}$A$_{g}$  & 258   & 260 \\
	\end{tabular}
\end{ruledtabular}
\end{table}

\begin{figure}[htb!]
	\centering
	\includegraphics[width=\columnwidth]{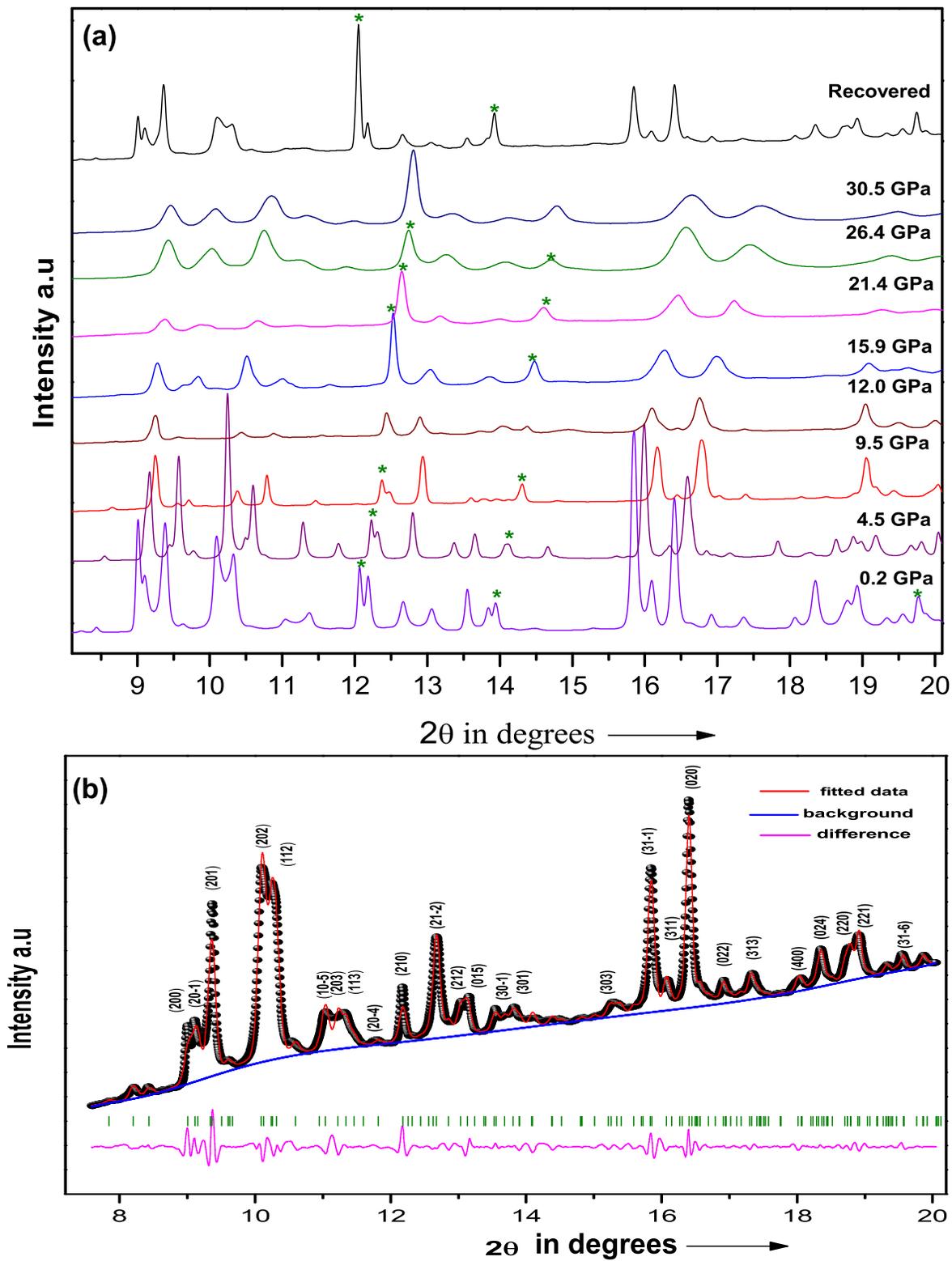}
	\caption{\label{Fig.1}(a) pressure evolution of XRD patterns at selected pressure points and (b) Lebail profile fitting of XRD pattern at ambient pressure. }
\end{figure} 
\begin{figure}[htb!]
	\centering
	\includegraphics[width=\columnwidth, height=20cm]{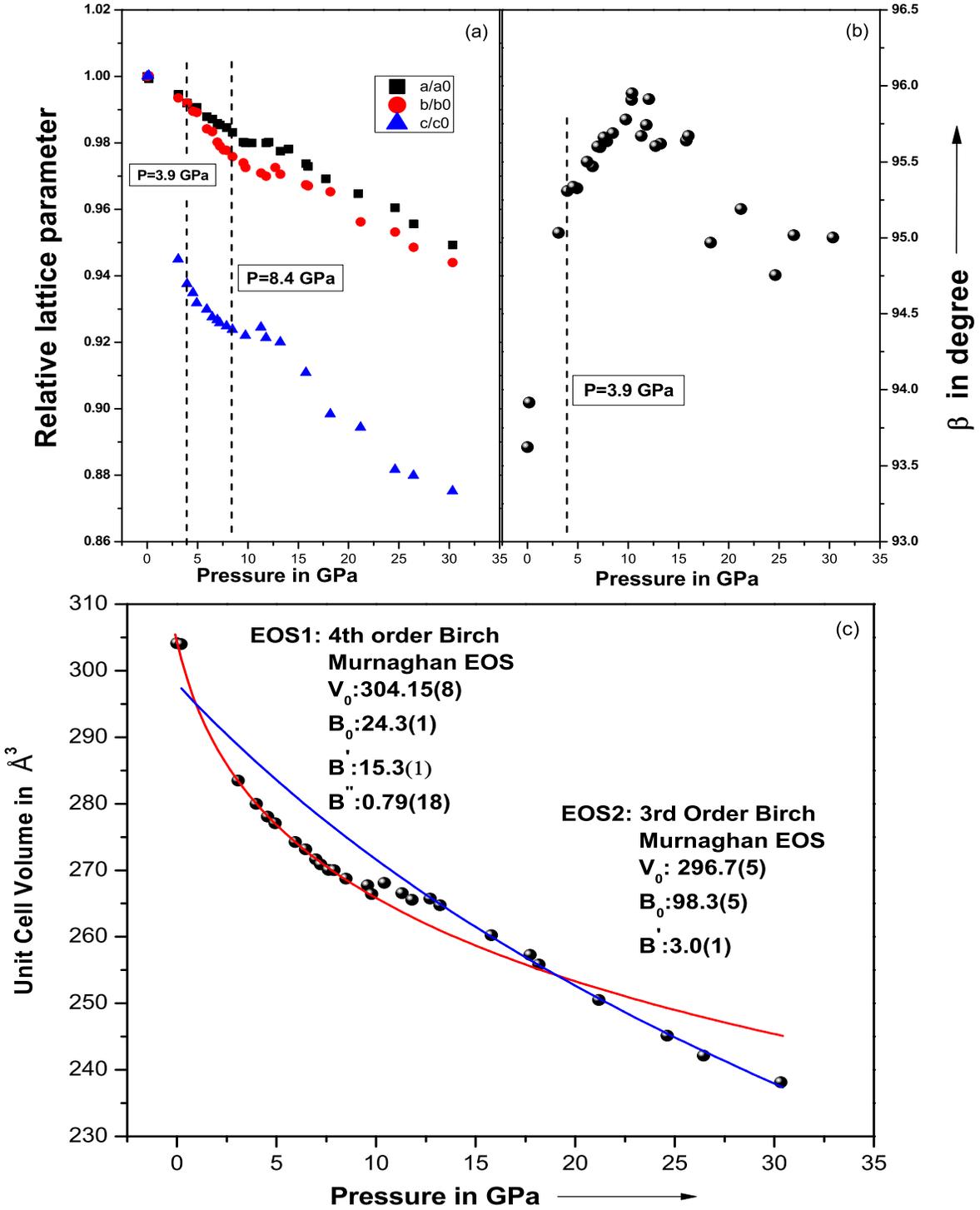}
	\caption{\label{Fig.2}(a) variation of $a/a_{0}$,$b/b_{0}$ and $c/c_{0}$ with pressure. The dashed vertical lines indicate pressure points showing the anomalous changes in lattice parameters. (b) Variation of lattice angle $\beta$ with pressure. (c) The variation of unit cell volume with applied pressure. The lines through data points show the EOS fits.}
\end{figure}

\begin{figure}[hbt!]
	\begin{center}
		\includegraphics[width=\columnwidth]{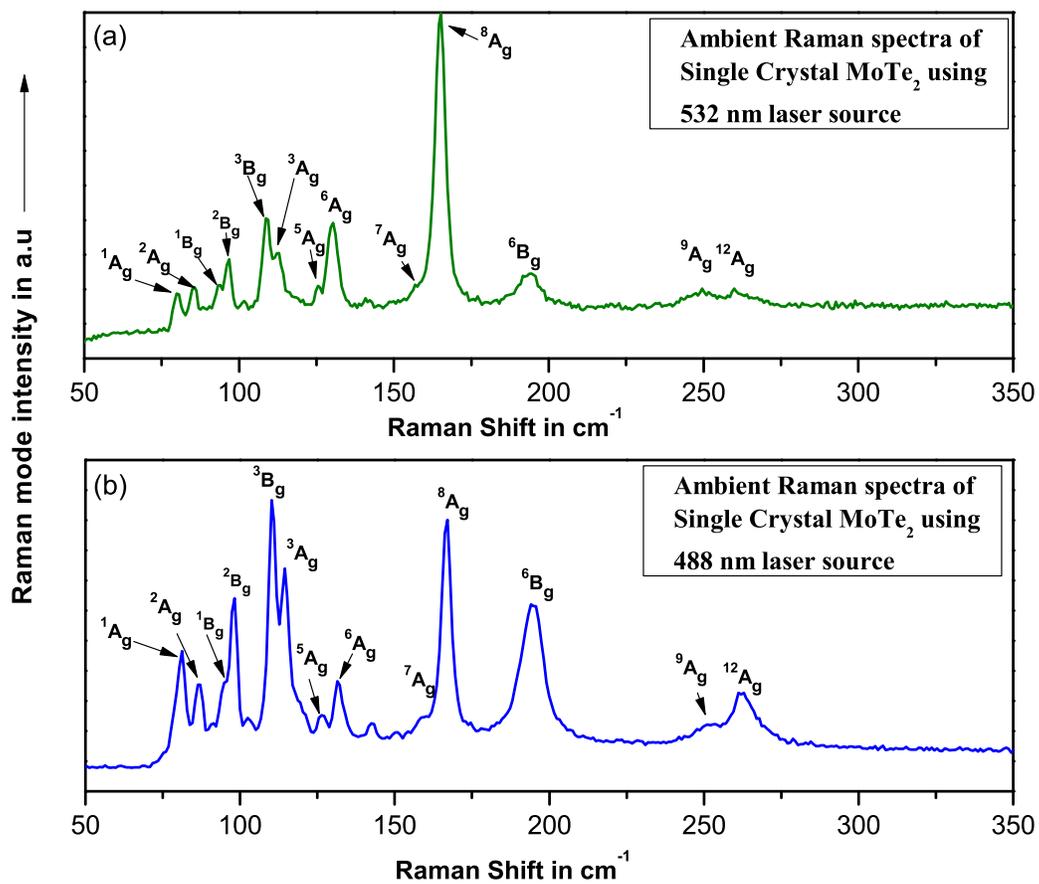}
	\end{center}
	\caption{\label{Fig.3}(Colour online) Ambient Raman Spectra using 532nm and 488nm excitation source. All the modes detected are shown in figure.}
\end{figure}

\begin{figure}[htb!]
	\centering
	\includegraphics[width=\columnwidth]{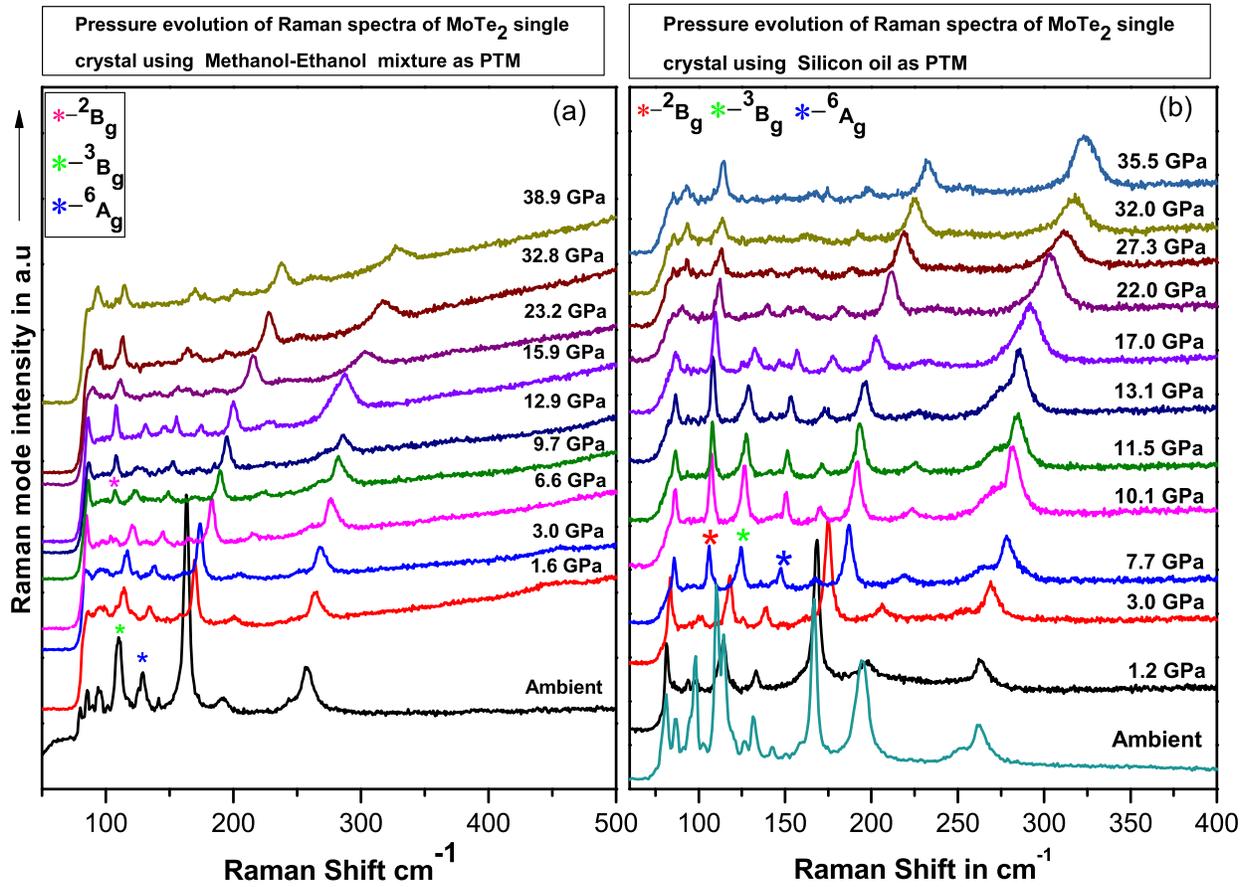}
	\caption{\label{Fig.4}(a) Pressure evolution of Raman Spectra using Methanol-Ethanol mixture as PTM and 532nm laser as excitation source.  (b) Pressure evolution of Raman Spectra using Silicone oil as PTM and 488nm laser as excitation source.} 
\end{figure}
\begin{figure}[htb!]
	\centering
	\includegraphics[width=\columnwidth]{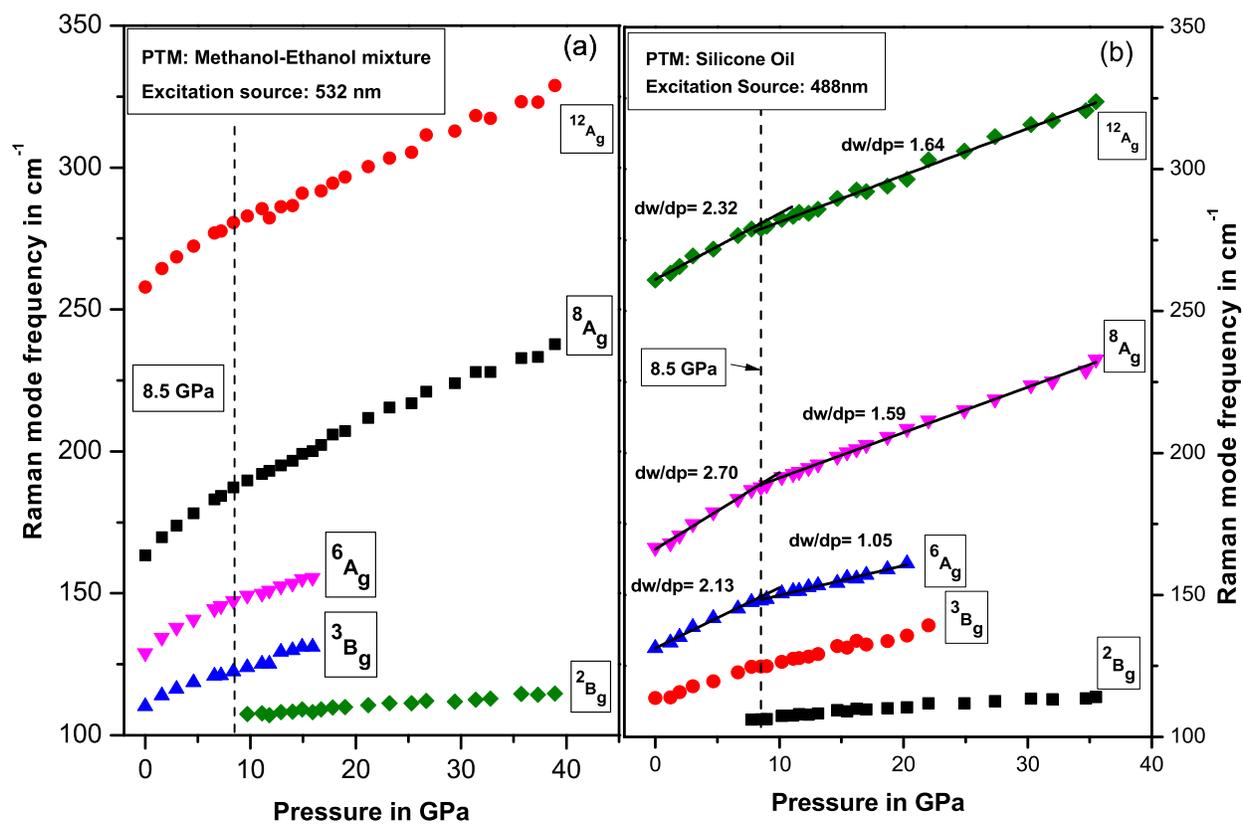}
	\caption{\label{Fig.5}Variation of few selected Raman mode frequencies with pressure, using (a) Methanol-Ethanol mixture as PTM and (b) Silicone oil as PTM. Around 8.5 GPa a change in slope is observed for most of the modes. }
\end{figure}[htb!]

\begin{figure}[htb!]
	\centering
	\includegraphics[width=\columnwidth]{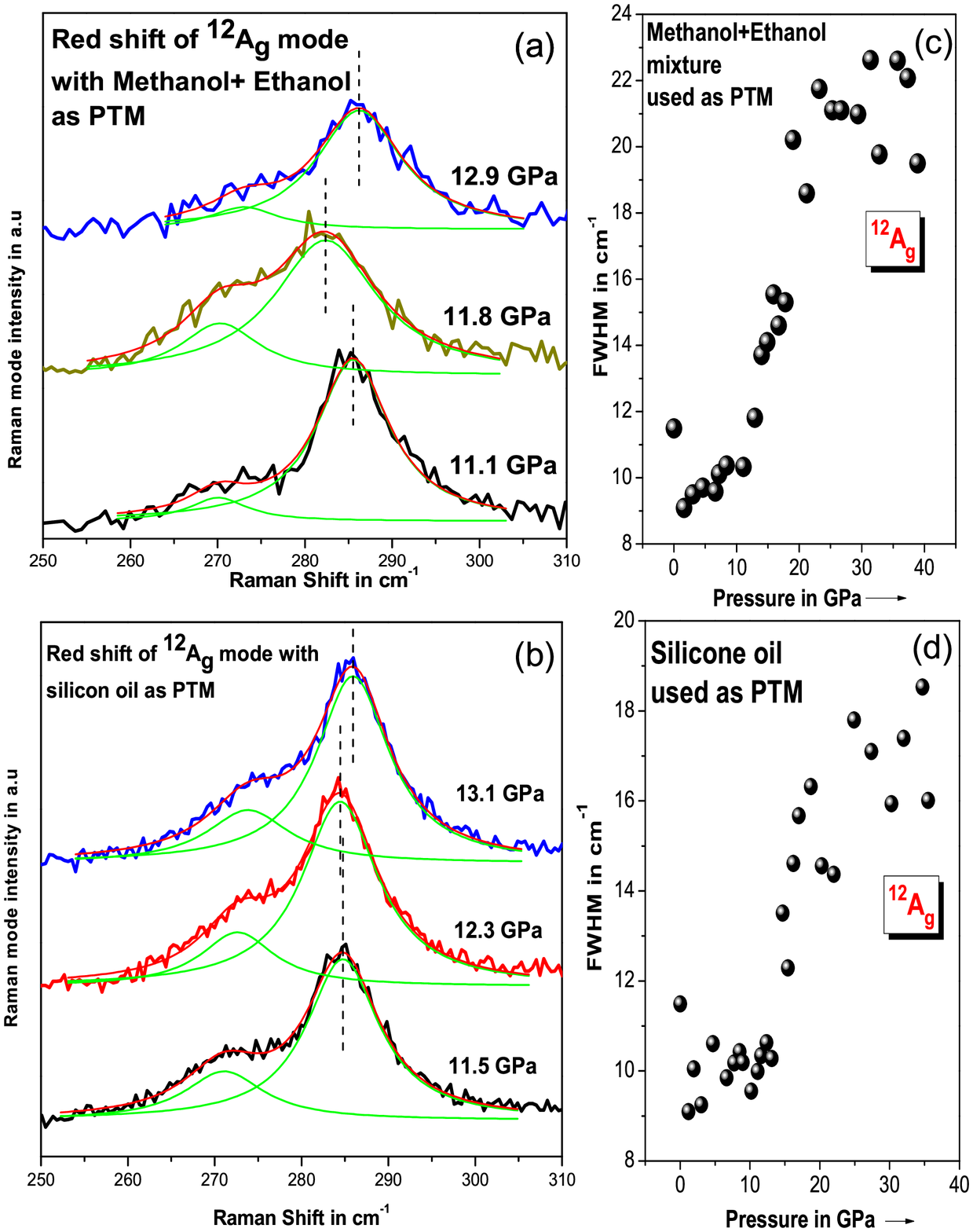}
	\caption{\label{Fig.6} Redshift of $^{12}$A$_{g}$ mode using (a) Methanol-Ethanol mixture as PTM, (b) Silicone oil as PTM. Pressure variation of FWHM for $^{12}$A$_{g}$ mode using (c)Methanol-Ethanol mixture as PTM, (d) Silicone oil as PTM.}
\end{figure}

%\begin{figure}[htb!]
%	\centering
%	\includegraphics[width=\columnwidth]{vestaPloDdiffPpoints.jpg}
%	\caption{\label{Fig.7} Redshift of $^{12}$A$_{g}$ mode using (a) Methanol-Ethanol mixture as PTM, %(b) Silicone oil as PTM. Pressure variation of FWHM for $^{12}$A$_{g}$ mode using (c)Methanol-Ethanol %mixture as PTM, (d) Silicone oil as PTM.}
%\end{figure}


\section{References}
\begin{references}
\bibitem{Qingjun}
Q.Song, H.Wang, X.Pan, X.Xu, Y.Li, F.Song, X.Wan, Y.Ye, and L.Dai, Scientific Report {\bf 7}, 1758 (2017).

\bibitem{Xiaoli}
X.Ma, P.Guo, C.Yi, Q.Yu, A.Zhang, J.Ji, Y.Tian, F.Jin, Y.Wang, K.Liu, T.Xia, Y.Shi, and Q.Zhang, Physical Review B {\bf 94}, 214105 (2016).

\bibitem{Yanpeng}
Y.Qi, P.G.Naumov, M.N.Ali, C.R.Rajamathi, W.Schnelle, O.Barkalov, M.Hanfland, S.-C. Wu, Chandra Shekhar, Y.Sun, V.Su, M.Schmidt, U.Schwarz, E.Pippel, P.Werner, R.Hillebrand, T.Forster, E.Kampert, S.Parkin, R.J.Cava, C.Felser, B.Yan, and S.A.Medvedev, Nature Communications {\bf 7}, 11038 (2016).

\bibitem{Pinku}
P.Saha, B.Ghosh, R.Jana, and G.D.Mukherjee, Journal of Applied Physics  {\bf 123}, 204306 (2018).

\bibitem{Pinku2}
P.Saha, B.Ghosh, A.Mazumder, and G.D.Mukherjee, preprint arXiv:1908. 10625 August 2019.

\bibitem{Zhao}
X.Zhao, H.Liu, A.F.Goncharov, Z.Zhao, V.V.Struzhkin, H.Mao, A.G.Gavriliuk, and X.Chen, Phys. Rev. B  {\bf 99}, 024111 (2019).

\bibitem{Mao}
H.K.Mao, J.Xu, and P.M.Bell, J. Geophys. Res. {\bf91}, 4673 (1986).

\bibitem{Preschera} 
C.Preschera, and V.B. Prakapenkaa, High Pressure Research: An International Journal {\bf35:3}, 223 (2015).

\bibitem{Dewaele}
A.Dewaele, M.Torrent, P.Loubeyre, and M.Mezouar, Phys. Rev. B {\bf78}, 104102 (2008).


\bibitem{Shirley}
R.Shirley, The CRYSFIRE 2002 System for Automatic Powder Indexing: Users Manual (TheLattice Press, Guildford, 2002).

\bibitem{Toby}
B.H.Toby, J. Appl. Crystallogr. {\bf34}, 210 (2001).

\bibitem{Dawson}
W.G.Dawson, and D.W.Bullett, J. Phys. C: Solid State Phys. {\bf 20}, 61594174 (1987).

\bibitem{Ahmad}
J.F.Ahmad, and I.Y.Alkammash, Journal of the Association of Arab Universities for Basic and Applied Sciences {\bf 12}, 17-22 (2012).

\bibitem{Angel}
R.J.Angel, J.G.-Platas, and M.Alvaro, Zeitschrift für Kristallographie {\bf 229}, 405 (2014).

\bibitem{Klotz}
S.Klotz, J-C.Chervin, P.Munsch, and G.L.Marchandh,  Journal of Physics D: Applied Physics{\bf 42}, 075413 (2009).
\end{references}
\end{document}